# Correcting motion induced fluorescence artifacts in two-channel neural imaging

Matthew S. Creamer[1], Kevin S. Chen[1], Andrew M. Leifer[1,2,*], Jonathan W. Pillow[1,3,*]

1. Princeton Neuroscience Institute, Princeton University
2. Department of Physics, Princeton University
3. Department of Psychology, Princeton University
* Correspondence: leifer@princeton.edu, pillow@princeton.edu

Imaging neural activity in a behaving animal presents unique challenges in part because motion from an animal's movement creates artifacts in fluorescence intensity time-series that are difficult to distinguish from neural signals of interest. One approach to mitigating these artifacts is to image two channels simultaneously; one that captures an activity-dependent fluorophore, such as GCaMP, and another that captures an activity-independent fluorophore such as RFP. Because the activity-independent channel contains the same motion artifacts as the activity-dependent channel, but no neural signals, the two together can be used to identify and remove the artifacts. Existing approaches for this correction, such as taking the ratio of the two channels, do not account for channel independent noise in the measured fluorescence. Moreover, no systematic comparison has been made of existing approaches that use two-channel signals. Here, we present Two-channel Motion Artifact Correction (TMAC), a method which seeks to remove artifacts by specifying a generative model of the fluorescence of the two channels as a function of motion artifact, neural activity, and noise. We use Bayesian inference to infer latent neural activity under this model, thus eliminating the motion artifact present in the measured fluorescence traces. We further present a novel method for evaluating ground-truth performance of motion correction algorithms by comparing the decodability of behavior from two types of neural recordings; a recording that had both an activity-dependent fluorophore (GCaMP and RFP) and a recording where both fluorophores were activity-independent (GFP and RFP). A successful motion-correction method should decode behavior from the first type of recording, but not the second. We use this metric to systematically compare five models for removing motion artifacts from fluorescent time traces. We decode locomotion from a GCaMP expressing animal 15x more accurately on average than from control when using TMAC inferred activity and outperform all other methods of motion correction tested.

## Introduction

Population fluorescent imaging of calcium-sensitive fluorescent indicators is a powerful approach for recording cellular neural dynamics [1,2]. Calcium imaging's widespread adoption has benefited from extensive development of computational algorithms to find and segment neurons of interest [3], to extract and denoise calcium signals [4,5], and to infer their underlying voltage signals [6]. An important goal of systems neuroscience is to probe the neural basis of animal behavior [7,8] and calcium imaging has been used to measure neural activity in awake and behaving animals [9–11]. However, animal motion during calcium imaging poses unique challenges both for segmenting and tracking neurons and for accurately extracting calcium traces. Many computational approaches have been proposed to register, segment and track neurons in the presence of motion [12–22]. These approaches account for the gross movement of a neuron relative to its neighbors or within the field of view. By contrast, there has been relatively few efforts to account and correct for motion-related changes to the extracted fluorescence intensity time-series itself. Accounting for these latter effects of motion, however, is particularly important because motion related changes to fluorescence can appear similar to behavior-related neural signals of interest. Here, we construct a method Two-channel Motion Artifact Correction (TMAC) which uses the fluorescence traces from two-channel imaging to correct for shared motion artifacts between the two channels.

When a neuron deforms or moves relative to an imaging plane, its fluorescence can change unrelated to neural activity **(Fig 1A)**. While multiple factors could contribute to these motion induced changes, at least some of them arise from subtleties of image acquisition and segmentation. Even though the number of fluorescent molecules in a neuron is constant, the neurons' shape, orientation and position are not. Accurately measuring fluorescence therefore requires carefully accounting for which voxels a neuron occupies, the intensity of each voxel, and the shape of the imaging hardware's point-spread function used to calculate the voxel. For example, as a neuron changes its shape with respect to the point spread function, it may cover less voxels but be brighter in the voxels that remain. In real-world imaging conditions it can be challenging to detect subtle changes to the boundary of a neuron, especially in densely labeled neural populations, and similarly the point-spread function is not always sufficiently characterized. These challenges are particularly acute in recordings of moving animals, such as *C. elegans* [11,23,24], *Hydra* [25], *Drosophila* larvae [26,27], and zebrafish larvae [28,29] that all exhibit large head deformations during movement.

Because these artifacts arise from motion, they have the potential to confound interpretations of the neural correlates of behavior. For example, if a neuron appears to change fluorescence during head bending, it may be hard to disambiguate whether that fluorescent transient reflects calcium activity or motion artifact. Two-channel imaging offers one strategy to account for motion-induced changes in fluorescence by measuring a calcium indicator like GCaMP in one channel and a calcium independent indicator like RFP in another. Any fluctuations of the calcium-independent channel must be either noise or motion artifact. In principle, this knowledge can be used to account for and correct artifacts in the calcium imaging channel. Crucially, whatever the source of these artifacts, as long as they affect the red and green channel equally, two-channel motion correction should be able to account for and remove them.

Two channel calcium imaging has its origins in the use of FRET based calcium indicators that use a donor and receptor and reports calcium signals as the ratio between the two channels [30]. The ratiometric approach further proved useful in freely moving animals [10], because the ratio is less sensitive to motion-induced fluorescent changes common to both channels. Single-channel GCaMP based indicators have now surpassed FRET-based indicators in terms of popularity because of their speed and brightness [31]. However, the use of two-channel imaging has persisted in moving animals in part for its ability remove motion artifacts with methods such as taking the ratio of the two channels [24,32–35], or by performing linear regression [11,36]. To date there has not been a systematic comparison of different mathematical approaches to account for motion induced changes to fluorescent intensity in two-channel imaging. The TMAC explicitly models the statistical distributions of the noise, artifact, and cell activity and uses Bayesian inference to find activity uncontaminated by motion artifact. We use experimental data to compare TMAC with 4 other approaches to motion artifact removal and demonstrate that it outperforms previous methods.

## Results
### Motion artifacts

We inspected two-channel calcium imaging recordings of neurons in moving *C. elegans* from Hallinen et al., 2021 (**Table S1**) for signs of motion-related calcium transients. We considered two types of recordings: those of control animals that express only the calcium-insensitive fluorophores GFP and RFP and those of calcium-sensitive animals that expressed the calcium indicator GCaMP in addition to RFP. In each case GFP or GCaMP was imaged in the green channel and RFP was imaged in the red channel. In order to account for differences in intensity, we divide each channel by its time average and report fluorescence as fold change from the mean. In control animals, the green and red fluorescent intensity fluctuated many times larger than the mean (**Fig. 1B**) and were highly correlated to each other (**Fig. 1D**). Because these animals contained no neural-related signals and only motion, we conclude that motion artifacts contribute to fluorescence variability. We also note the strong correlation between the GFP and RFP activity suggests that the motion artifact is shared between the two channels. These shared fluctuations are what make motion correction using a second channel possible.

In GCaMP recordings, we also observed a correlation between the green and red channels (**Fig 1DE)**, although less so than in control recordings. The presence of activity-related signal in the green channel likely explains this difference. The correlation that remains between the red and the green channel suggests that the red channel could be used to correct for shared motion artifacts in the green channel.

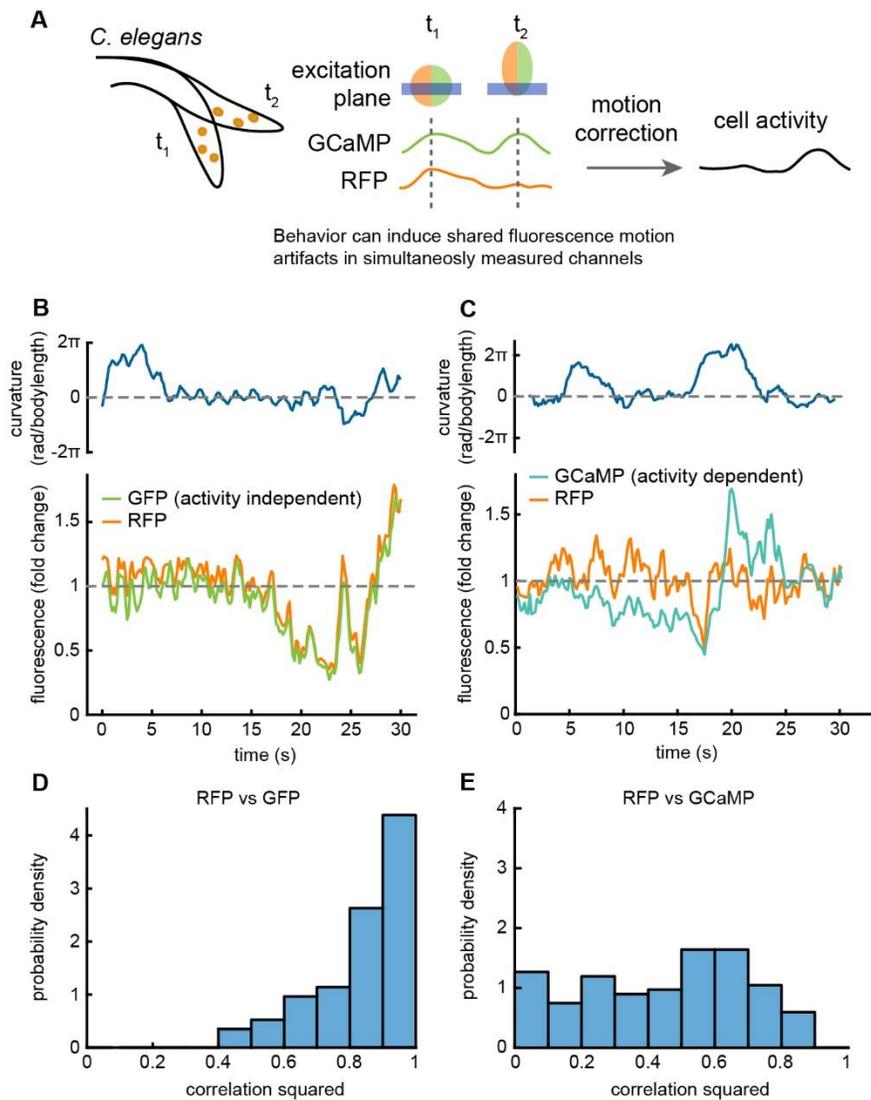

**Figure 1. Behavior induces motion artifacts in optical recordings.** A) Diagram of a moving *C. elegans* undergoing motion and deformation during simultaneous measurements of GCaMP and RFP fluorescence. RFP intensity fluctuations reflect motion or noise. GCaMP intensity fluctuations reflect both motion, calcium activity, and noise. B) *Top*, Animal body curvature. *Bottom* fluorescence of a neuron from a whole-brain recording of a moving worm expressing both GFP and RFP. Both fluorophores are activity independent, yet we observe large highly correlated fluctuations in the two-channel fluorescence. C) *Top*, Body curvature. *Bottom*, fluorescence of a single neuron taken from a whole-brain recording in a moving worm expressing both GCaMP and RFP. The two channels still show correlated fluctuations despite the activity dependence of GCaMP. D) Histogram of the Pearson correlation coefficient squared between the red and the green channel for a dataset of whole-brain recordings from 10 moving GFP, RFP control worms. E) Same as D but in 9 GCaMP, RFP worms.

**Model**

To account for fluctuations in the fluorescence induced by motion artifact, we propose a method called Two-channel Motion Artifact Correction (TMAC) (**Fig. 2A**). We constructed a latent variable model of calcium fluorescence and then use this generative model to infer the latent activity of the neuron. We first process the fluorescence data by dividing by the time average, such that the processed data has mean 1 and units of fold change from the mean. By specifying our model fluorescence as fold change from the mean the parameters do not depend strongly on experimental conditions such as cell expression level, laser intensity, or image acquisition time, allowing us to compare across different neurons and fluorophores (**Fig. 1BC**).

We define the fluorescence from a neural-activity dependent green channel and a neural-activity independent red channel like those described above as follows. Let *r* and *g* denote vectors of the preprocessed red channel and green channel fluorescence from the same neuron. We assume that *r* is the sum of a latent motion artifact time series *m* and additive Gaussian white noise $\varepsilon_r$. The green channel measurements **g**, arise as the sum of the same latent motion artifact *m* and a latent time series of neural activity *a*, plus Gaussian white noise $\varepsilon_g$. To obtain a complete generative model, we assume Gaussian process priors over *m* and *a*, which specifies that they evolve smoothly over time. Formally the model can be written as:

$$r = 1 + m + \epsilon_r$$
$$g = a + m + \epsilon_g$$
$$m \sim N(0, \Sigma_m)$$
$$a \sim N(1, \Sigma_a)$$
$$\epsilon_{r,g} \sim N(0, \sigma^2_{r,g} I)$$

Where N denotes a Gaussian distribution and $\sigma^2_r$ and $\sigma^2_g$ denote the variance of the additive Gaussian noise in red and green channels, respectively. The latent time series of the motion artifact *m* and neural activity *a* are each defined by a Gaussian process with covariance $\Sigma_a$ and $\Sigma_m$. The prior covariances $\Sigma_a$ and $\Sigma_m$ are each parameterized by a pair of hyperparameters, ($\sigma^2_a$, $\tau_a$) and ($\sigma^2_m$, $\tau_m$), where $\sigma^2$ denotes the prior variance or amplitude, and $\tau$ denotes the length scale, controlling the degree of smoothness. The covariances are defined as radial basis functions, such that the covariance between two time points depends only on the squared distance between them:

$$\Sigma_{m_{(t,t')}} = \sigma^2_m exp\left(-\frac{(t-t')^2}{2\tau^2_m}\right)$$
$$\Sigma_{a_{(t,t')}} = \sigma^2_a exp\left(-\frac{(t-t')^2}{2\tau^2_a}\right)$$

Given this model structure, data *r* and *g*, parameters *a* and *m*, and hyperparameters $\theta$ ($\sigma^2_a$, $\tau_a$, $\sigma^2_m$, $\tau_m$, $\sigma^2_r$, $\sigma^2_g$), we fit the model in two steps. We first define the marginal likelihood $p(r,g|\theta)$ which is used to infer the hyperparameters $\hat{\theta}$ and we then define the posterior $p(a,m|r,g,\hat{\theta})$ given the learned hyperparameters which we use to infer *a* and *m*. The marginal likelihood is the joint probability of the data *r* and *g* and parameters *a* and *m* integrated over *a* and *m*.

$$p(r,g|\theta) = \int p(r,g,a,m|\theta) da dm$$
$$p(r,g|\theta) = N\left(\begin{bmatrix}1\\1\end{bmatrix}, \begin{bmatrix}\Sigma_m + \sigma^2_r I & \Sigma_m \\ \Sigma_m & \Sigma_a + \Sigma_m + \sigma^2_g I\end{bmatrix}\right)$$
$$\hat{\theta} = argmax_\theta\, p(r,g|\theta)$$

The marginal probability is the likelihood function for the hyperparameters. Since it does not depend on *a* and *m*, it can optimized with respect to the hyperparameters without overfitting to any particular instantiation of *a* or *m*.

By Bayes Rule the posterior p(*a,m*|*r,g*,θ) is proportional to the product of the likelihood of *a* and *m* p(*r,g*|*a,m*,θ) and the prior p(*a,m*|θ). The posterior is the probability of the *a* and *m* given the data *r*, *g*, the maximum of which is the maximum a posteriori (MAP) estimates $\hat{a}$ and $\hat{m}$.

$$p(a,m|r,g,\hat{\theta}) \propto p(r,g|a,m,\hat{\theta})p(a,m,\hat{\theta}) = N\left(\begin{bmatrix}1+m\\a+m\end{bmatrix}, \begin{bmatrix}\sigma_r^2 I & 0\\0 & \sigma_g^2 I\end{bmatrix}\right) N\left(\begin{bmatrix}1\\0\end{bmatrix}, \begin{bmatrix}\Sigma_a & 0\\0 & \Sigma_m\end{bmatrix}\right)$$

$$\hat{a},\hat{m} = argmax_{a,m}\, p(a,m|r,g,\hat{\theta})$$

**Model validation**

In order to demonstrate that our model correctly infers ground truth activity and variances, we used TMAC to generate a synthetic dataset (**Fig. 2B** top) with 5000 time points and a correlation time scale which roughly matches experimental datasets from **Fig. 1**, such that 6 time points approximately corresponds to one second. We then used TMAC to infer both the activity and motion artifact from this synthetic dataset (**Fig 2B** middle and bottom). We find a good correspondence between the inferred activity from TMAC and the true activity (**Fig. 2C**). The model also accurately infers the hyperparameters of the model (**Fig. 2D**).

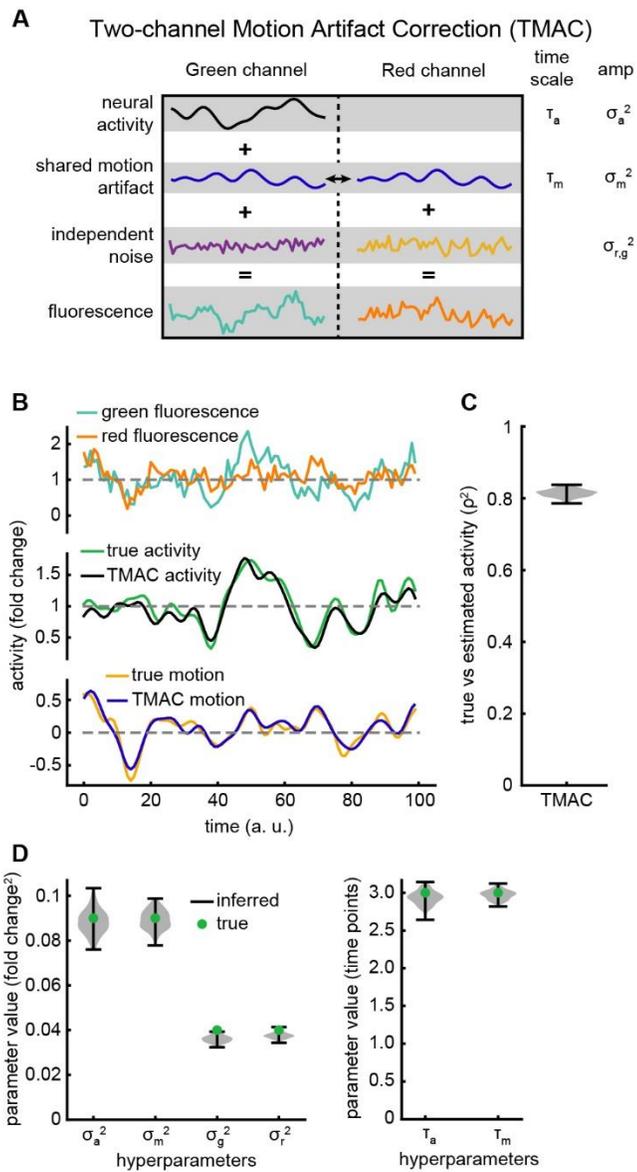

**Figure 2**. **TMAC can infer activity and hyperparameters from synthetic data.** A) Diagram of the structure of TMAC. The green channel is modeled as the sum of the calcium activity, motion artifact, and independent gaussian noise. The red channel is modeled as the sum of the motion artifact and independent gaussian noise. The motion artifact is shared between the two channels. B) Top: fluorescence from a synthetic green and red channel. Middle and bottom: inferred activity and motion compared with the true activity and motion. C) Correlation squared between estimated activity from TMAC and true activity over many synthetic datasets. D) Violin plot of inferred and true parameter values when fitting TMAC.

**Decoding Behavior**

We next sought to evaluate TMAC on experimentally acquired neural population calcium imaging recordings in moving worms from Hallinen et al. 2021. The learned hyperparameters provide a convenient method to characterize the recordings. We identified a putative high signal-to-noise neuron, by finding a neuron that the model associated with a high activity variance ($\sigma^2_a$) and low motion and noise variances ($\sigma^2_{m,r,g}$) (**Fig. 3A** middle). We also identified a putative low signal-to-noise neuron by finding a neuron that the model associated with high motion artifact ($\sigma^2_m$) and low activity and noise ($\sigma^2_{a,r,g}$) (**Fig. 3A** bottom). In the case where TMAC estimates high signal-to-noise, TMAC predicts that a smoothed version of the green channel represents true calcium activity (**Fig. 3A** middle). In the case where TMAC estimates that most fluorescent fluctuations are due to motion, TMAC deviates from the fluorescent fluctuations and returns a flatter inferred activity (**Fig. 3A** bottom).

We next wanted to evaluate TMAC's performance on removing motion artifacts. Since we lack access to the ground truth calcium activity, we considered the problem of decoding animal curvature from neural activity (**Fig. 3B**). Using decodability as a metric of performance is challenging because motion artifacts may also provide information about the target behavior. In some regimes, successfully removing motion artifacts could in principle reduce decoding performance rather than improve it.

We therefore evaluated motion correction performance by inspecting the decodability of two types of recordings; one that had both activity-dependent and activity-independent fluorophores (GCaMP and RFP), and a control worm that had two activity-independent fluorophores (GFP and RFP), both from Hallinen et al., 2021. Although the GFP control worm contains no neural signal, the worm's behavior can still be decoded, albeit poorly, by relying on behavior-related information in the motion artifact. We reasoned that a successful motion correction algorithm should reduce the decodability of motion artifacts while retaining the ability to decode from neural signals. We therefore define a new metric that evaluates motion correction performance as the ratio of its decodability from recordings of GCaMP animals to its decodability of recordings of GFP animals (**Fig. 3C**). Using this metric, we compare four models used for motion correction in two channel imaging. Single channel GCaMP which is just the green channel fluorescence with no motion correction, the ratio model discussed above, a linear regression model [11], an ICA based approach [37], and TMAC. The linear regression model finds the best linear fit of the red channel to the green channel and then subtracts off that best fit from the green channel. The ICA approach performs ICA using the red and green channel as inputs and returns the independent component least correlated to the red fluorescence.

The activity fit by TMAC is ~15x more effective at predicting curvature from a GCaMP expressing worm than a control worm (**Fig. 3C**). We calculated the correlation between the activity inferred by TMAC and the red channel and they did not correlate strongly, suggesting that the model is successful at removing motion artifacts common to the two channels (**Fig. 3DE**).

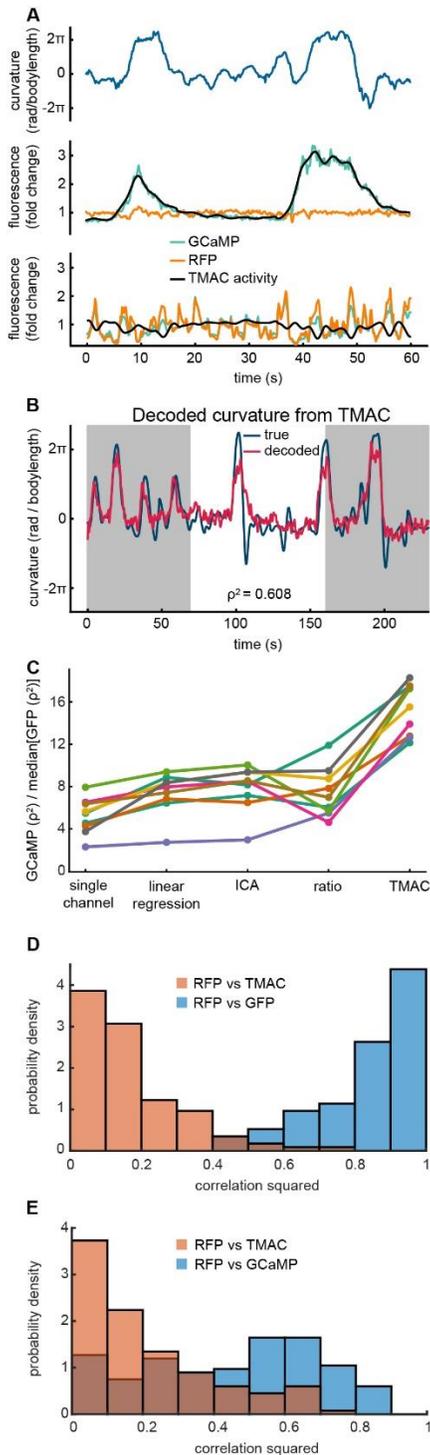

**Figure 3. TMAC reduces decodable motion artifacts in experimental data.** A) *Top*, animal body curvature over time. *Middle*, GCaMP and RFP fluorescence from a neuron that TMAC estimates to have high signal to noise, recorded from a behaving worm. *Bottom*, GCaMP and RFP fluorescence from a different neuron in that same recording that TMAC estimates to have large motion artifacts. B) Time trace of animal curvature and predicted behavior, decoded from activity inferred by TMAC in a GCaMP worm. Gray shaded regions were used to train the decoder, white region was held out and used to evaluate decoding performance. C) Ratio of decoding accuracy ($\rho^2$) when decoding GCaMP divided by the median accuracy for a GFP worm. D) Histogram over all neurons of correlation squared between RFP and activity inferred by TMAC from a GFP worm. RFP vs GFP data the same as in **Fig. 1DE**. E) Same as F but in a GCaMP worm.

**Discussion**

Motion artifacts are prevalent in recordings of behaving animals (**Fig. 1BC**). These artifacts can appear as signals of interest, reducing the interpretability of the data. Here we presented TMAC, a model which infers the latent neural activity without these artifacts by leveraging information in two-channel calcium imaging recordings. We demonstrate that TMAC substantially reduces decodable motion artifacts in experimental data (**Fig 3A**) and outperforms five alternatives (**Fig 3C**).

The model we propose uses an additive interaction between the fluorescence transients caused by motion artifact and the fluorescence transients caused by neural activity. It is unclear in experimental data what the true interaction between artifact and activity is. However, even if the true interaction is not additive, TMAC will perform well under model mismatch because it can be thought of as a linear approximation of the true interaction between *a* and *m*. Consider the case where the activity and motion interact multiplicatively. Multiplicative interactions have linear components when the two variables have nonzero means. The linear component of the multiplicative interaction is clear by writing the equation for the green channel with no noise.

$$g = (a+1)(m+1) = am + a + m + 1$$

We can be sure that the *a* and *m* have a mean (in this case 1), because fluorescent data is nonnegative. Furthermore, the *am* term itself can be approximated by *a* + *m* as long as *a* and *m* are small. If the approximation of *a* and *m* as Gaussian is valid, we can be sure that the *am* and higher order terms are small, because the standard deviations of *a* and *m* must be small relative to 1 to avoid negative values. For these reasons, while TMAC is a linear approximation to the true interaction between *a* and *m*, it is an approximation that is highly accurate. An interesting avenue for future work would be to consider asymmetric distributions which can retain both high variability and positivity and may better approximate the true activity distribution.

In this work we demonstrated TMAC's ability to remove decodable motion artifacts from calcium induced fluorescence. However, TMAC will remove motion artifacts in any type of two-channel imaging, so long as one channel is activity-independent and both channels share the same motion artifact component. TMAC could therefore be applied to a wide range of two-channel imaging modalities including for voltage imaging (Bando et al., 2019), fiber photometry when using an isosbestic wavelength [39,40], or two-channel two-photon imaging [41].

# Supplementary information

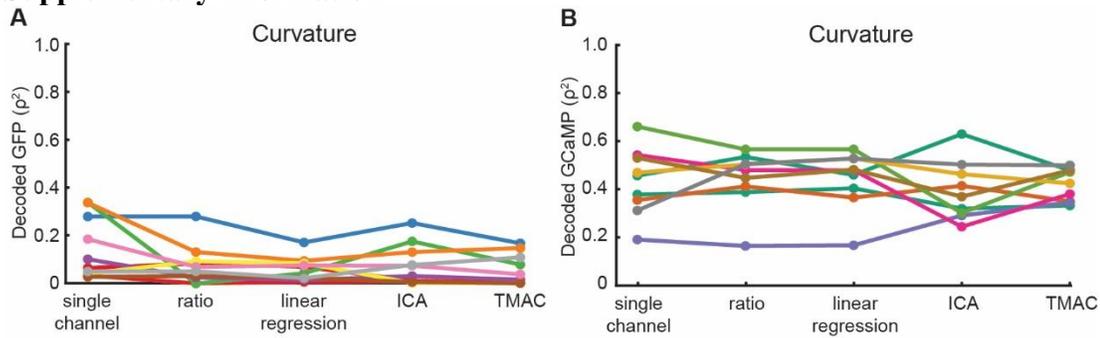

**Figure S1. Motion artifact correction for GCaMP and GFP expressing animals**. A) Decoding accuracy when decoding whole-body curvature from GFP recordings with different motion correction methods applied. These animals do not express an activity dependent fluorophore so all decoding comes from motion artifacts. B) Decoding accuracy when decoding whole-body curvature from GCaMP expressing animals. The ratio of B to the median for each method in A is the value reported in **Fig. 3C**.

**Table S1. Data source for each figure.**

|  | **Dataset from Hallinen et al. 2021** |
|---|---|
| **Fig 1B** | AML18_A |
| **Fig 1C** | AML310_A |
| **Fig 1D** | AML18_A-C, E-K |
| **Fig 1E** | AML310AC, AML32A-G |
| **Fig 3AB** | AML310_A |
| **Fig 3C** | blue: AML310_A<br>orange: AML310_C<br>red: AML32_A<br>yellow: AML32_B<br>green: AML32_C<br>brown: AML32_D<br>purple: AML32_E<br>gray: AML32_F<br>pink: AML32_G<br><br>AML18_A-C, E-K (Used to calculate median[GFP decoding]) |
| **Fig 3D** | AML18_A-C, E-K |
| **Fig 3E** | AML310_AC, AML32_A-G |
| **Fig S1A** | blue: AML18_A<br>teal: AML18_B<br>green: AML18_C<br>pink: AML18_E<br>purple: AML18_F<br>orange: AML18_G<br>red: AML18_H<br>brown: AML18_I<br>gray: AML18_J<br>yellow: AML18_K |
| **Fig S1B** | blue: AML310_A<br>orange: AML310_C<br>red: AML32_A<br>yellow: AML32_B<br>green: AML32_C<br>brown: AML32_D<br>purple: AML32_E<br>gray: AML32_F<br>pink: AML32_G |